\documentclass[a4paper,fleqn,usenatbib]{mnras}

\usepackage[T1]{fontenc}
\usepackage{ae,aecompl}

\usepackage{graphicx}	% Including figure files
\usepackage{amsmath}	% Advanced maths commands
\usepackage{amssymb}	% Extra maths symbols
\usepackage{mathptmx}
\usepackage{txfonts}

\title[67P Aggregate Acceleration]{Acceleration of Individual, Decimetre-sized Aggregates in the Lower Coma of Comet 67P/Churyumov-Gerasimenko}

\author[J. Agarwal et al.]{Jessica Agarwal,$^{1}$\thanks{E-mail: agarwal@mps.mpg.de (JA)}
M. F. A'Hearn,$^{2}$
J.-B. Vincent,$^{1}$
C. G\"uttler,$^{1}$
S. H\"ofner,$^{1}$
H. Sierks,$^{1}$
\newauthor 
C. Tubiana,$^{1}$
C. Barbieri,$^{3}$
P. L. Lamy,$^{5}$
R. Rodrigo,$^{8,9}$ 
D. Koschny,$^{4}$
H. Rickman,$^{6,7}$ 
\newauthor 
M. A. Barucci,$^{10}$
J.-L. Bertaux,$^{11}$
I. Bertini,$^{12}$
S. Boudreault,$^{1}$
G. Cremonese,$^{13}$
\newauthor 
V. Da Deppo,$^{14}$
B. Davidsson,$^{15}$
S. Debei,$^{16}$
M. De Cecco,$^{17}$
J. Deller,$^{1}$
S. Fornasier,$^{10}$
\newauthor 
M. Fulle,$^{18}$
A. Gicquel,$^{1}$
O. Groussin,$^{19}$
P. J. Guti\'errez,$^{20}$
M. Hofmann,$^{1}$
S. F. Hviid,$^{21}$
\newauthor 
W.-H. Ip,$^{22}$
L. Jorda,$^{19}$
H. U. Keller,$^{23}$
J. Knollenberg,$^{21}$
J.-R. Kramm,$^{1}$
E. K\"uhrt,$^{21}$
\newauthor 
M. K\"uppers,$^{24}$
L. M. Lara,$^{20}$
M. Lazzarin,$^{3}$
J. J. Lopez Moreno,$^{20}$
F. Marzari,$^{3}$
\newauthor 
G. Naletto,$^{12,14,25}$ 
N. Oklay,$^{1}$
X. Shi$^{1}$
and N. Thomas$^{26}$
\\
\begin{minipage}{\textwidth}
\vspace{\baselineskip}
$^{1}$Max-Planck Instit\"ut fur Sonnensystemforschung, Justus-von-Liebig-Weg 3, 37077 G\"ottingen, Germany; 
$^{2}$Department for Astronomy, University of Maryland, College Park, MD 20742-2421, USA; 
$^{3}$Department of Physics and Astronomy "G. Galilei", University of Padova, Vic. Osservatorio 3, 35122 Padova, Italy; $^{4}$Research and Scientific Support Department, European Space Agency, 2201 Noordwijk, The Netherlands; 
$^{5}$Laboratoire d'Astrophysique de Marseille, UMR 7326 CNRS \& Aix-Marseille Universit\'e, 38 rue Fr\'{e}d\'{e}ric Joliot-Curie, 13388 Marseille cedex 13, France; 
$^{6}$Department of Physics and Astronomy, Uppsala University, Box 516, 75120 Uppsala, Sweden; 
$^{7}$PAS Space Research Center, Bartycka 18A, 00716 Warszawa, Poland; 
$^{8}$Centro de Astrobiologia (INTA-CSIC), European Space Agency, European Space Astronomy Centre (ESAC), P.O. Box 78, E-28691 Villanueva de la Canada, Madrid, Spain; 
$^{9}$International Space Science Institute, Hallerstrasse 6, 3012 Bern, Switzerland; 
$^{10}$LESIA, Observatoire de Paris, CNRS, UPMC Univ Paris 06, Univ. Paris-Diderot, 5 Place J. Janssen, 92195 Meudon Pricipal Cedex, France; 
$^{11}$LATMOS, CNRS/UVSQ/IPSL, 11 Boulevard d'Alembert, 78280 Guyancourt, France; 
$^{12}$Centro di Ateneo di Studi ed Attivit{\`a} Spaziali "Giuseppe Colombo" (CISAS), University of Padova, Via Venezia 15, 35131 Padova, Italy; 
$^{13}$INAF Osservatorio Astronomico di Padova, Vicolo dell'Osservatorio 5, 35122 Padova, Italy; 
$^{14}$CNR-IFN UOS Padova LUXOR, Via Trasea 7, 35131 Padova, Italy; 
$^{15}$Jet Propulsion Laboratory, M/S 183-301, 4800 Oak Grove Drive, Pasadena, CA 91109, U.S.A.; 
$^{16}$Department of Industrial Engineering University of Padova Via Venezia, 1, 35131 Padova, Italy; 
$^{17}$University of Trento, via Sommarive, 9, Trento, Italy; 
$^{18}$INAF - Osservatorio Astronomico di Trieste, via Tiepolo 11, 34143 Trieste, Italy; 
$^{19}$Aix Marseille Universit\'e, CNRS, LAM (Laboratoire d'Astro-physique de Marseille) UMR 7326, 13388, Marseille, France; 
$^{20}$Instituto de Astrofisica de Andalucia-CSIC, Glorieta de la Astronomia, 18008 Granada, Spain; 
$^{21}$Institute of Planetary Research, DLR, Rutherfordstrasse 2, 12489 Berlin, Germany; 
$^{22}$Institute for Space Science, National Central University, 32054 Chung-Li, Taiwan; 
$^{23}$Institute for Geophysics and Extraterrestrial Physics, TU Braunschweig, 38106 Braunschweig, Germany; 
$^{24}$ESA/ESAC, PO Box 78, 28691 Villanueva de la Ca\~nada, Spain; 
$^{25}$Department of Information Engineering, University of Padova, Via Gradenigo 6/B, 35131 Padova, Italy; 
$^{26}$Physikalisches Institut, Sidlerstrasse 5, University of Bern, CH-3012 Bern, Switzerland
\end{minipage}
}

\date{Accepted XXX. Received YYY; in original form ZZZ}

\pubyear{2016}

\begin{document}
\label{firstpage}
\pagerange{\pageref{firstpage}--\pageref{lastpage}}
\maketitle

\begin{abstract}
We present OSIRIS/NAC observations of decimetre-sized, likely ice-containing aggregates ejected from a confined region on the surface of comet 67P/Churyumov-Gerasimenko. The images were obtained in January 2016 when the comet was at 2 AU from the Sun out-bound from perihelion. We measure the acceleration of individual aggregates through a two-hour image series. Approximately 50\% of the aggregates are accelerated away from the nucleus, and 50\% towards it, and likewise towards either horizontal direction. The accelerations are up to one order of magnitude stronger than local gravity, and are most simply explained by the combined effect of gas drag accelerating all aggregates upwards, and the recoil force from asymmetric outgassing, either from rotating aggregates with randomly oriented spin axes and sufficient thermal inertia to shift the temperature maximum away from an aggregate's subsolar region, or from aggregates with variable ice content. At least 10\% of the aggregates will escape the gravity field of the nucleus and feed the comet's debris trail, while others may fall back to the surface and contribute to the deposits covering parts of the northern hemisphere. The rocket force plays a crucial role in pushing these aggregates back towards the surface. Our observations show the future back fall material in the process of ejection, and provide the first direct measurement of the acceleration of aggregates in the innermost coma ($<$2km) of a comet, where gas drag is still significant.  
\end{abstract}

\begin{keywords}
comets:general -- comets: individual: 67P/Churyumov-Gerasimenko -- zodiacal dust
\end{keywords}

%%%%%%%%%%%%%%%%%%%%%%%%%%%%%%%%%%%%%%%%%%%%%%%%%%

%%%%%%%%%%%%%%%%% BODY OF PAPER %%%%%%%%%%%%%%%%%%

\section{Introduction}
Since the time when comet 67P/Churyumov-Gerasimenko (hereafter, 67P) was selected as the target of the Rosetta mission in 2003, evidence has been growing that the dust ejected from this comet was dominated by particles that show little sensitivity to radiation pressure, typically of order 100 $\mu$m and larger. Comet 67P has a debris trail that consists of such large particles ejected over the last 8 perihelion passages after the comet's close encounter with Jupiter in 1959 \citep{sykes-lebofsky1986, ishiguro2008, kelley-reach2008}. Smaller particles are strongly affected by solar radiation pressure that blows them away from the immediate vicinity of the comet on timescales of weeks, while the large particles disperse only slighly along the comet's orbital path, forming the debris trail. To feed sufficient material into the trail to account for its observed brightness, the large trail particles must have dominated the optical cross section also in the coma around the perihelion passage \citep{agarwal-mueller2010, soja-sommer2015}.
The comet was expected to eject chunks of up to 1 m in size \citep{fulle-colangeli2010}. A significant abundance of large dust grains was also inferred from the polarisation properties of light scattered in the coma of 67P \citep{hadamcik-sen2010}, and from the low expansion velocities of the developing coma when the comet returned from aphelion \citep{tozzi-patriarchi2011}. 
The prediction of significant quantities of large grains deviated from the established picture that dust scattering in cometary comae was dominated by micron-sized grains. 

When Rosetta arrived at 67P in August 2014, even at the heliocentric distance of 3.6 AU, individual large particles of up to 1 m in size were abundant in the coma, some of which may have been orbiting the comet since its previous perihelion passage \citep{rotundi-sierks2015, fulle-marzari2016}.
Throughout the duration of the Rosetta mission at 67P, images obtained with the Optical, Spectroscopic, and Infrared Remote Imaging System (OSIRIS, \citet{keller-barbieri2007}) showed significant quantities of individual grains that are clearly distinguishable from the diffuse coma background. While a comprehensive and quantitative analysis of this phenomenon is not yet available, qualitatively this appearance of the coma is consistent with an unexpectedly high abundance of large grains combined with a low abundance of smaller grains, enhancing the contrast between the individual, large grains and the diffuse background.

Independently, large terrains on the cometary surface appear to be covered in material that has fallen back to the surface from the coma \citep{thomas-sierks2015,thomas-davidsson2015}, implying that a significant quantity of refractory material is lifted from the surface but does not leave the gravity field of the nucleus. 

We present observations of decimetre-sized aggregates lifted from a confined region on the surface of comet 67P. We measure their projected velocities and accelerations, and find that they likely contain ice, the sublimation of which influences the aggregates' motion in a decisive way. We expect that part of the aggregates will enter the comet's debris trail, while part will fall back to the surface or enter bound orbits. We see the material at the large end of the dust size distribution in the process of ejection.
To our knowledge, the data discussed here show for the first time the acceleration of individual dust aggregates near the surface of a comet, where gas drag is still active. Aggregates of comparable size have been observed in the coma of comet Hartley 2 to be accelerated by the the rocket effect from sublimating ice \citep{kelley-lindler2013, kelley-lindler2015}, but those aggregates were observed further away from the comet where their motion was already decoupled from the gas flow.

\citet{gundlach-blum2015} predict that the decimetre-size range provides optimum conditions for particles to be lifted from the surface due to the trade-off between low van-der-Waals force between neighbouring aggregates and comparatively low weight. This prediction coincides with the size of the aggregates we present in this study.

In Section~\ref{sec:measurements}, we describe the data set and our measurement of the aggregates' projected velocity and acceleration, and infer their size from the measured brightness. In Section~\ref{sec:dynamics}, we constrain the source region, initial velocities, and future trajectories of the aggregates, followed in Section~\ref{sec:forces} by a discussion of the significance of gravity, rocket force, and gas drag for the motion of the aggregates, and of a scenario that explains the observed accelerations as the result of the combined forces of gas drag and rocket force.
In the concluding Section~\ref{sec:conclusions} we summarise our results and address open questions.

\section{Measurements}
\label{sec:measurements}
We analyse a set of images obtained on 6 January 2016 with the Narrow Angle Camera (NAC) of the OSIRIS camera system on board the Rosetta spacecraft. NAC has a 2048 x 2048 pixel CCD detector and a field of view (FOV) of 2.2$^\circ \times$ 2.2$^\circ$. The linear angular resolution is 18.6 $\mu$rad/pixel.
The images were obtained with the orange filter, centred at 649 nm with a bandwidth of 85 nm. The images were bias-subtracted, flat-fielded, corrected for angular distortion, and flux calibrated relative to standard stars using the OSIRIS calibration pipeline software \citep{tubiana-guettler2015}.

At the time of our observations, comet 67P was at a distance of 2.059 AU from the Sun on its way outward from perihelion, and 1.602 AU from the Earth. Rosetta was at a distance of 86.6 km from the comet. At this distance, a single NAC pixel corresponds to 1.6 m, and the full NAC FOV covers 3.3 km.
The phase angle between the Sun, comet centre, and Rosetta was 89.9$^\circ$. 

Between UT 07:01:03 and 08:51:15, 12 pairs of images were obtained at a 10 minutes' cadence. Each pair consisted of one short (0.24 s) and one long (6 s) exposure taken at 12 s from each other. Embedded between the first and second image pair was a {\it movie} sequence, comprising ten additional image pairs of 0.24 s exposure time and 6 s offset between images of a given pair. The pairs were taken at a 40 s cadence. 
%The observation and exposure times are listed in Table~\ref{tab:times}. 

\subsection{Aggregate Identification and Velocities}
\label{subsec:identify}
\begin{figure}
	\includegraphics[width=\columnwidth]{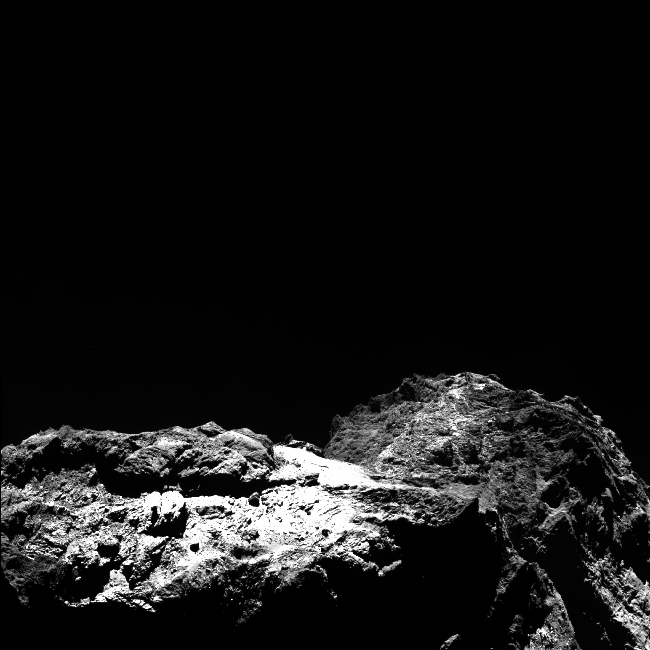}
    \caption{Single NAC image obtained on 6 January 2016 at UT 07:03:02.6 with 0.24 s exposure time in orange filter. The comet is at the bottom with the big lobe to the left and facing the observer, and the small lobe in the background in the right part of the image. The Sun was at the top of the image. The colour scale is linear from 0 (black) to 0.0007 W m$^{-2}$ nm$^{-1}$ sr$^{-1}$ (white).}
    \label{fig:single}
\end{figure}

\begin{figure}
  %\includegraphics[width=\columnwidth]{comb_max_fullFOV_2e-6-1e-5.png}
  \includegraphics[width=\columnwidth]{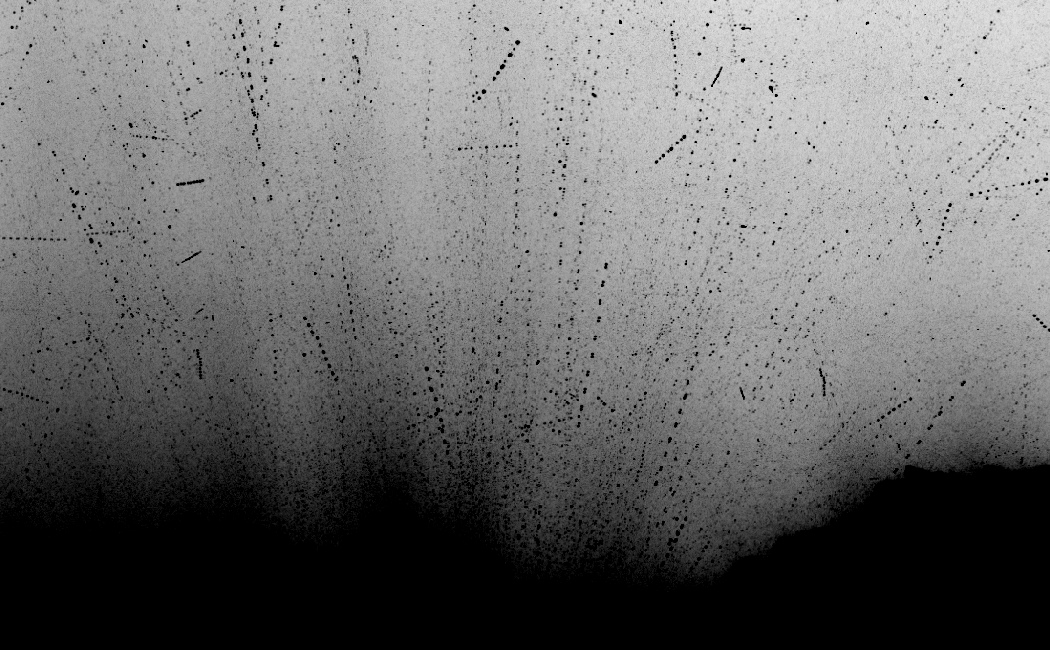}
    \caption{Maximum-stack of 20 images obtained between UT 07:03:03 and 07:09:09. The colour scale is inverted linear from 2 $\times$ 10$^{-6}$ (white) to 10$^{-5}$ W m$^{-2}$ nm$^{-1}$ sr$^{-1}$ (black). The displayed image section spans 1.17$^\circ$ x 0.73$^\circ$.
}
\label{fig:comb}
\end{figure}

During our observations, the camera was pointing at the coma above the subsolar limb of the nucleus. An example image is shown in Figure~\ref{fig:single}.
The comet is visible in the lower part of the images with the big lobe facing the observer and the small lobe behind it. The Sun is at the top of the images. Each image shows many point sources embedded in the diffuse coma. In the following, we identify these point sources as pieces of refractory material (henceforth called aggregates) and analyse their motion.
During the acquisition of the 6 min movie sequence, the pointing of the camera relative to the comet was stable. The projected position of the comet centre of mass moved by less than one NAC pixel, and the azimuth of the Sun direction changed by $<$0.04$^\circ$. In order to identify individual aggregates and measure their motion, we stacked all 20 movie images retaining the brightest value at each CCD position. In the stacked image (Figure~\ref{fig:comb}), moving grains are visible as tracks of up to 20 separate dots, depending on their projected speed and rotation state. 

\begin{figure}
  \includegraphics[width=\columnwidth]{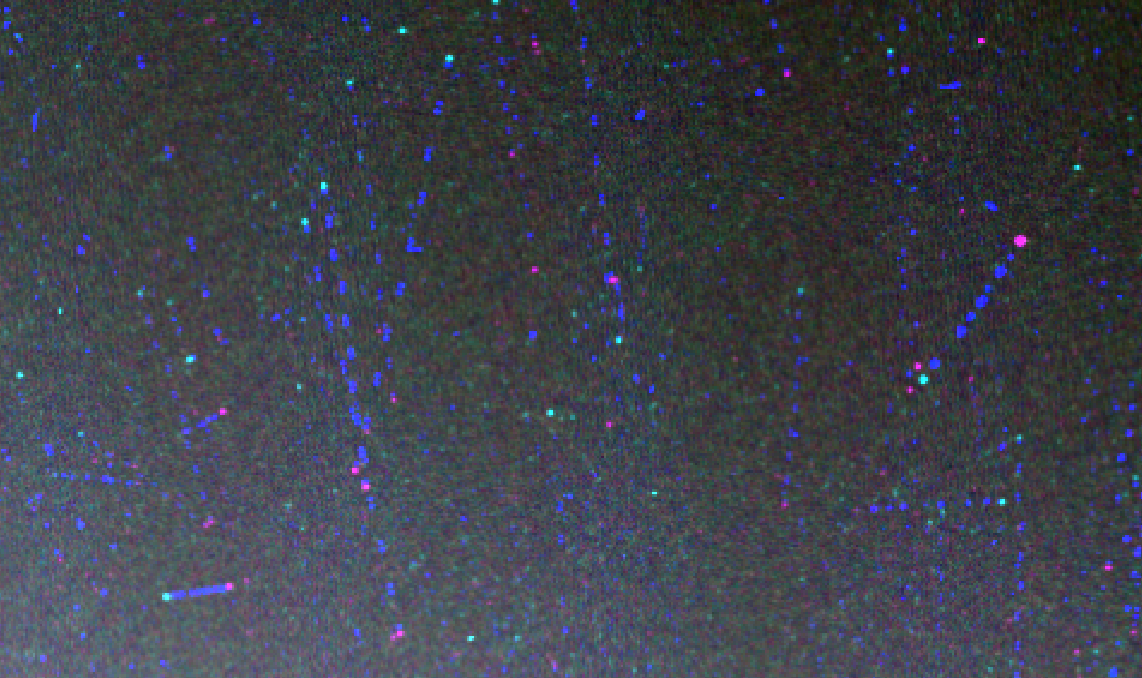}

  \vspace{1mm}
  \includegraphics[width=\columnwidth]{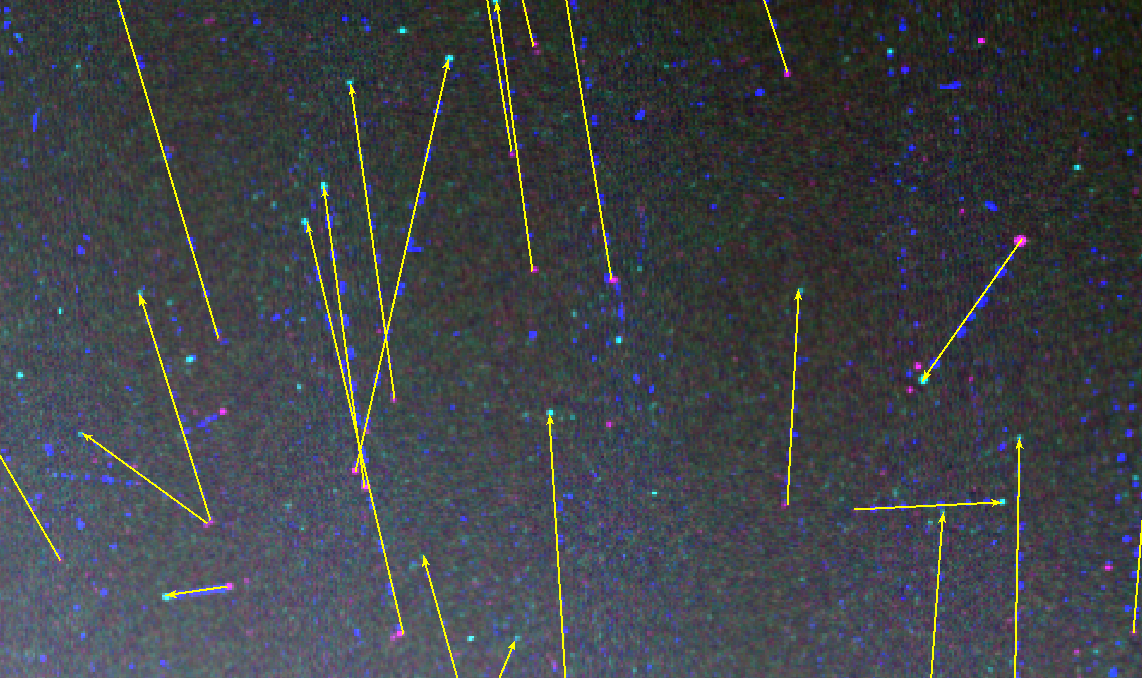}
    \caption{RGB composite of the 20 movie frames. The blue channel shows the maximum stack of all 20 movie frames (cf. Figure~\ref{fig:comb}). The red and green channels show the first and last image of the movie sequence, respectively. Both panels show the same image section and have an angular size of 0.5$^\circ \times$ 0.3$^\circ$. The lower panel in addition shows the identified tracks as yellow arrows.
}
\label{fig:rgb}
\end{figure}

To identify tracks of individual aggregates we displayed the movie images as an RGB composite: the blue channel represented the stack of 20 movie frames shown in Figure~\ref{fig:comb}, while the red and green channels contained the first and last of the movie frames, respectively. A zoomed detail of this RGB composite is shown in the upper panel of Figure~\ref{fig:rgb}.

We considered a track as identifiable if it was visible in the red and green channels and showed eight pairs of equidistant blue spots in between. We also accepted tracks where one or two of the blue spots were apparently missing but which still had the length of nine equidistant intervals. An example of such a track is the one going from top right to bottom left in the right part of Figure~\ref{fig:rgb}, where a double blue spot at position 8 is not visible. We attribute this non-detection to the rotation of an irregularly shaped aggregate, perhaps of a disc-like shape, that we caught edge-on at this particular movie station. \citet{fulle-ivanovski2015} also found that oblate grain shapes are well suited to model rotating, millimetre-sized grains detected with OSIRIS.

\begin{figure}
  \includegraphics[width=\columnwidth]{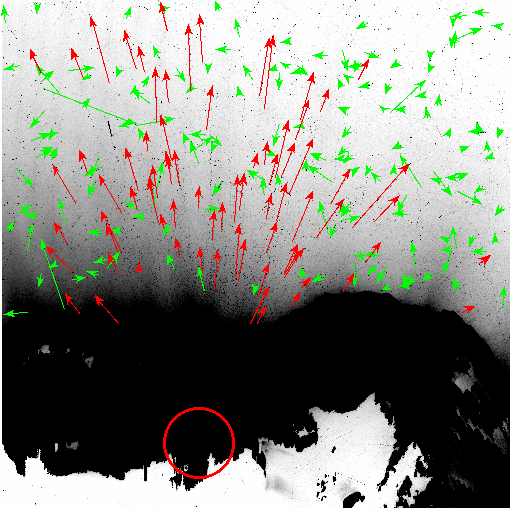}
    \caption{Tracks of moving aggregates identified in both the first and last image of the movie sequence. Red arrows indicate the group of radial tracks, while the randomly oriented tracks are marked green. The approximate radiant of the radial tracks is marked by a red circle.}
    \label{fig:tracks}
\end{figure}

For each of the 238 aggregates identified in this manner, we measured the position, length and direction of the track (Figures~\ref{fig:rgb}, bottom, and \ref{fig:tracks}). The tracks seem to divide in two groups: the group of {\it radial} tracks that seem to originate from a source region near the bottom centre of the image, and the {\it randomly oriented} tracks that are on average shorter than the radial tracks and have no preferred direction. To distinguish between the two groups, we define a track as radial if its direction deviates less than 15$^\circ$ from the direction given by a point at image coordinates (800,270), marked by the circle in Figure~\ref{fig:tracks}, to the footpoint of the track. We assume that the radial tracks represent aggregates from a specific region on the surface, and that the randomly oriented tracks correspond to particles closer to the camera. We concentrate in the following analysis on the 76 identified radial tracks, marked red in Figure~\ref{fig:tracks}. 
Assuming that the radial tracks are at the same distance from the camera as the comet, we convert the track length to the distance travelled during the 366 s duration of the movie sequence and to the projected velocity of the aggregates. The average track length is 145 px with a standard deviation of 60 px, which corresponds to projected velocities of (0.63 $\pm$ 0.26) m s$^{-1}$.

\subsection{Acceleration}
\label{subsec:acc}
\begin{figure}
  \includegraphics[width=\columnwidth]{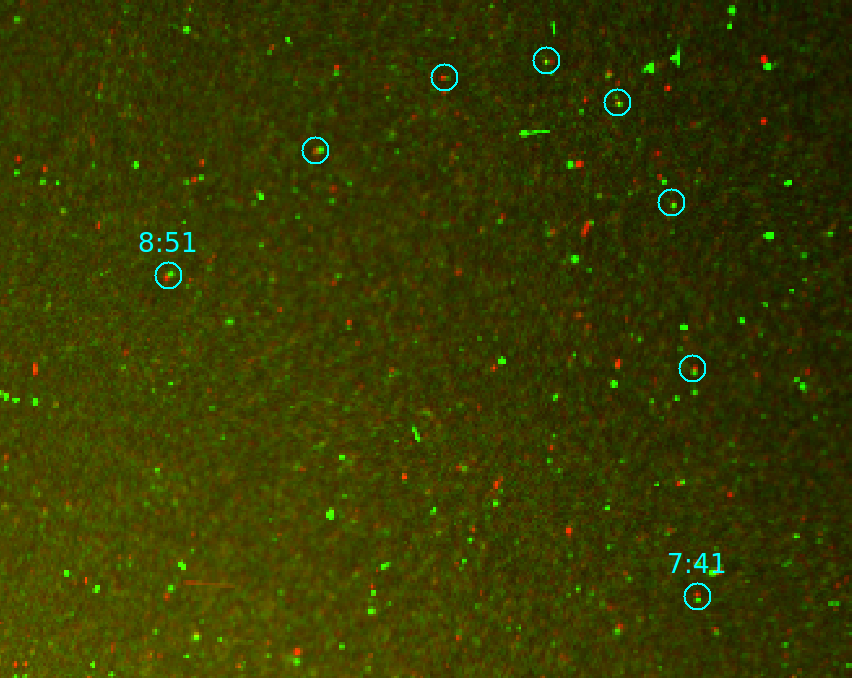}
    \caption{Trajectory of one aggregate falling back towards the nucleus in projection. The image is a red-green composite of two stacked images. The green channel shows the stack of all 12 short exposures taken at 10 minutes intervals, and the red channel is a stack of all corresponding long exposures, taken at 12 s offset. The time sequence shows that the grain was initially moving upward (green position below red one), then, both were superimposed, and later, the green position was above the red position, indicating downward motion.} %TBC: re-do this in blue-yellow.}
    \label{fig:invert}
\end{figure}

Starting from the stacked image of the movie sequence, we manually tracked the path of each radially moving aggregate in the consecutive image pairs taken at intervals of 10 minutes. 
The projected velocity measured in the stacked movie frames gave us a good estimate of the aggregate's position in the first post-movie image pair, obtained 2 minutes after the last movie frame. Having identified the aggregate at this station, we proceeded in a similar way through all subsequent image pairs, inferring the expected position at station $n$ based on the velocity measured between stations $n-1$ and $n-2$, until the aggregate either left the FOV or could not be detected for a different reason (possibly due to viewing it at an unfavourable rotational phase). With five exceptions, there was only one possible candidate near each expected position. The identifications were made in maximum stacks of the two images obtained close in time at each 10 minute station, such that these frames were far less crowded than the one shown in Figure~\ref{fig:invert}, which is a composite of 12 image pairs. The identification was further aided by the circumstance that the images were taken as pairs, such that the direction of motion of an aggregate at each station was known. Two thirds of the aggregates were detected in more than one post-movie frame.

\begin{figure}
  \includegraphics[width=\columnwidth]{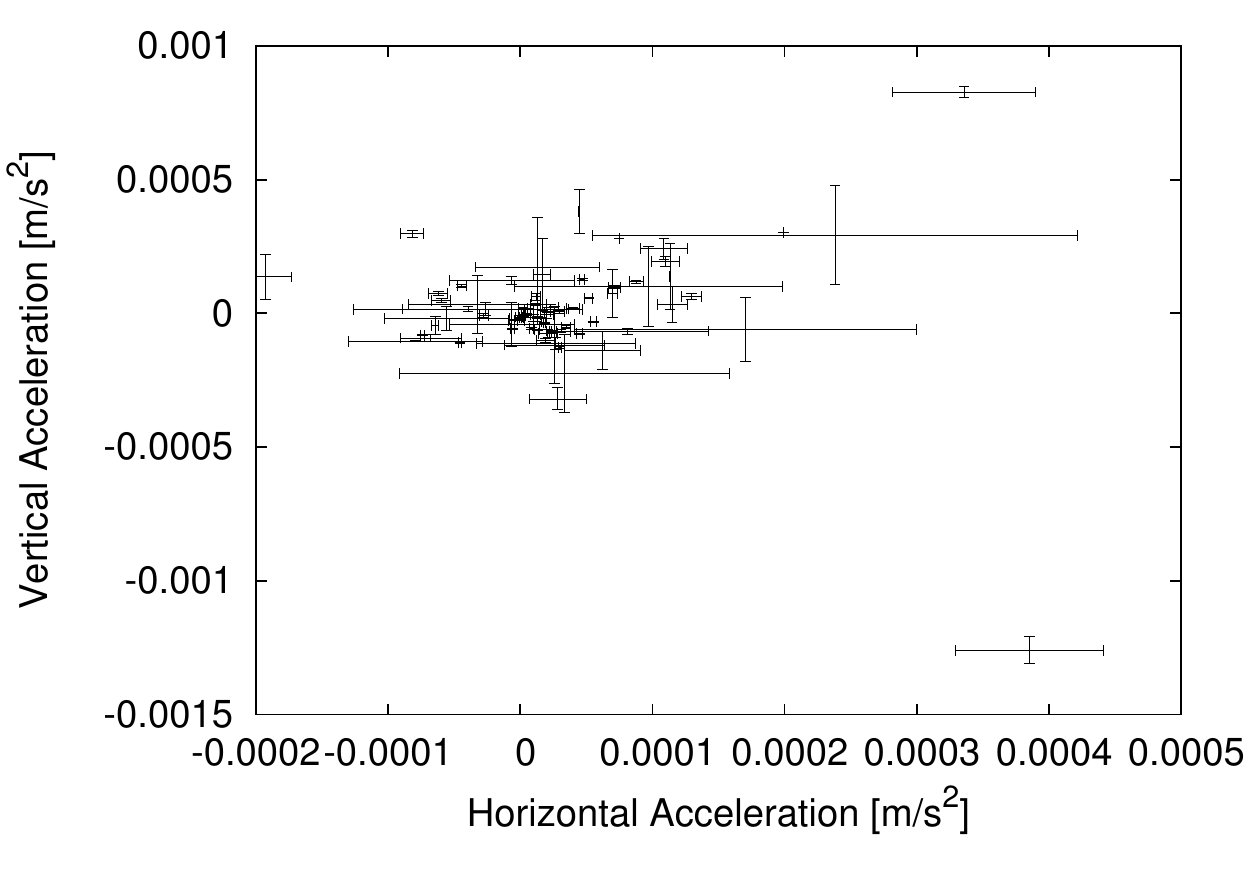}
    \caption{Projected acceleration in the image x- and y-direction of the 76 radially moving aggregates, derived from a linear fit to the measured projected velocities.}
    \label{fig:acc}
\end{figure}

We measured the positions of the radially moving aggregates in each image, and applied a minor correction for the small apparent motion of the projected centre of the comet (15 NAC pixels in the vertical direction over 2h).
From the positions, we derived the mean projected velocities for each time interval between consecutive image pairs. Generally, the velocities change linearly with time, which allowed us to derive a constant projected acceleration for each aggregate. The acceleration components along the image directions are shown in Figure~\ref{fig:acc}.
Approximately equal numbers of aggregates are accelerated along the positive (upward), and negative (downward) vertical direction. One aggregate is seen to invert its apparent direction of motion during the 2h interval of our observations (Figure~\ref{fig:invert}). Its acceleration falls within the typical range of accelerations and we found no indication that this aggregate is in any way special. Rather it is likely that we conincidentally observed this particular aggregate near the highest point of its trajectory, while other aggregates will have fallen back only after the end of our observing campaign.

\begin{figure}
  \includegraphics[width=\columnwidth]{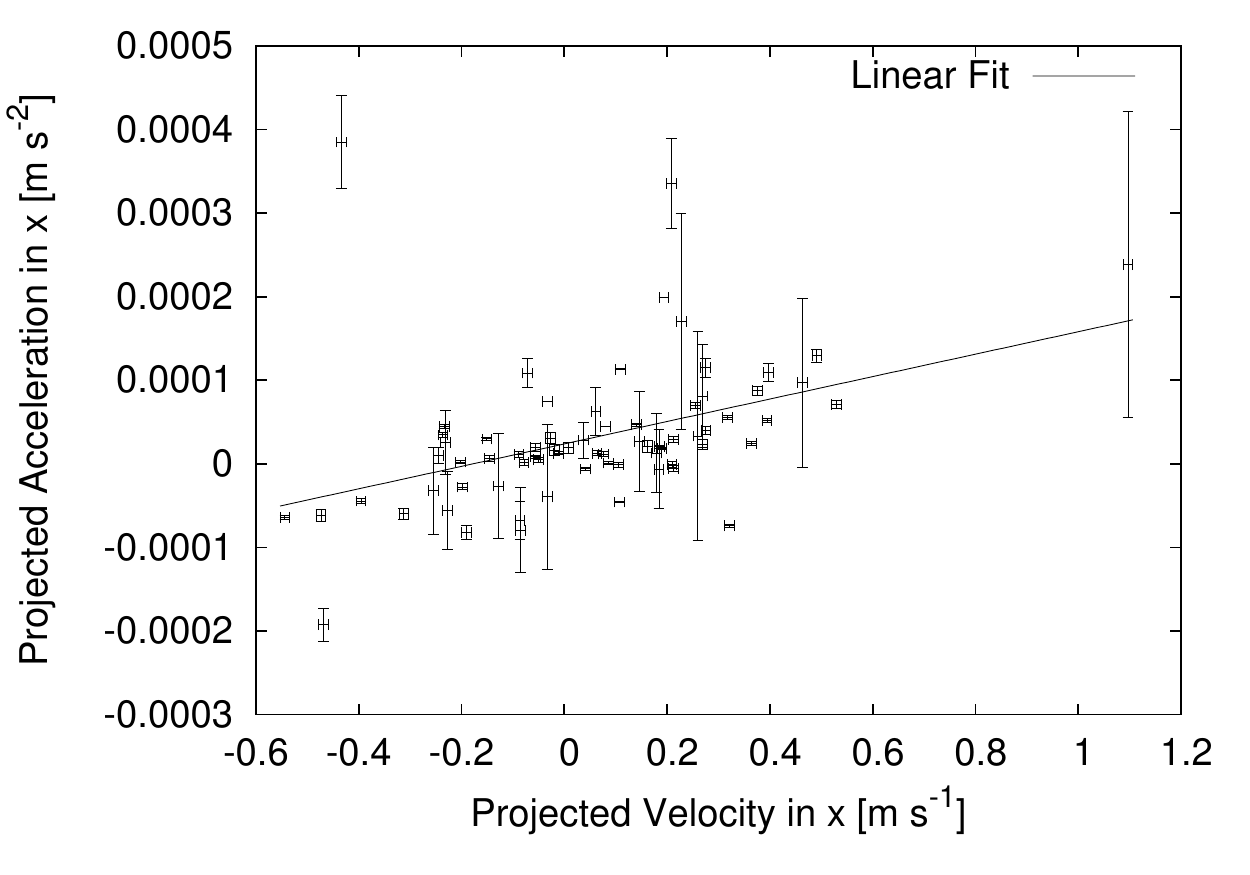}
  \includegraphics[width=\columnwidth]{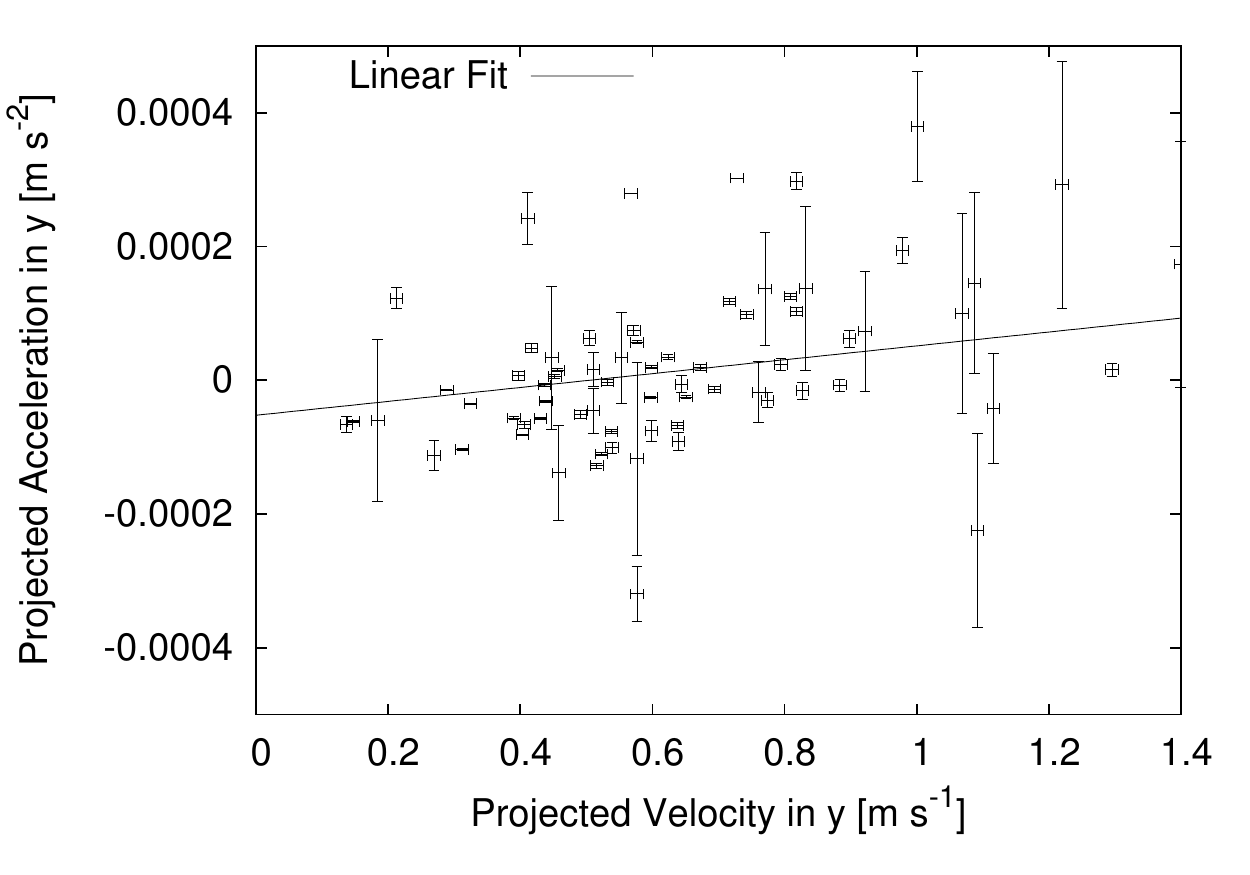}
    \caption{Correlation between acceleration and velocity, for the horizontal (top) and vertical (bottom) components. The solid lines represent linear fits with $a_x$ = 1.3 10$^{-4}$ s$^{-1} v_x$ + 0.2 10$^{-4}$ m s$^{-1}$ and $a_y$ = 10$^{-4}$ s$^{-1} v_y$ - 0.5 10$^{-4}$ m s$^{-1}$.}
    \label{fig:a_v}
\end{figure}

We observe a correlation between the acceleration and velocity for both the horizontal and the vertical component (Figure~\ref{fig:a_v}). This indicates a causal relationship between the on-going acceleration and the observed velocities, i.e. that similar accelerating forces have been acting in the past and that they had a significant influence on the motion of the aggregates. This makes an impulsive scenario unlikely where the aggregates received initial velocitites due to e.g. an explosion and subsequently were subject only to gravity. Forces that would exert a continuous acceleration include gas drag and rocket force due to the sublimation of ice embedded in the aggregates.
The time constants in Figure~\ref{fig:a_v} are of order 10$^{-4}$ s$^{-1}$. If the aggregates started from zero initial speed and the acceleration was constant during the time before our observations, this corresponds to a flight time of 2.8 h. We will discuss in Section~\ref{subsec:v0} that likely the time since ejection from the surface is much shorter and that therefore the aggregates received a significant fraction of their velocities during an early stage of their flight.

\subsection{Photometry}
\label{subsec:phot}
We measured the aggregate brightness in the individual images using the {\it Image Reduction and Analysis Facility} (IRAF). During the short exposures (0.24 s), even the fastest aggregates travelled only one sixth of a pixel, while during the long (6 s) exposures, they are trailed over up to 4 pixels. Comparing the radial profiles of some of the brightest aggregates in the short exposures to that of field stars, we found no indication of the aggregates being different from point sources. 
The radial profile of a field star 3$\times$ brighter than the brightest, and 10$\times$ brighter than the average aggregate, showed that the measured flux is independent of aperture size for aperture radii $>$ 5 pixels, and that an aperture with a radius of 3 pixels samples 96\% of the total flux. The radial profiles of aggregates (in the short exposures) and of a field star of comparable brightness indicate that at this lower flux level, an aperture radius of 3 pixels provides the optimum trade-off between sampling the PSF as completely as possible and limiting background noise. We therefore used circular apertures of 3 pixels radius for the short exposures, and of 7 pixels radius for the long exposures. 

The background was measured in a concentric annulus 3 pixels wide and separated from the inner aperture by 2 pixels, and subtracted. The measured fluxes typically show a temporal variability by up to a factor of two, but no obvious trend with time. 
%Part of the variability is due to rotation of an irregularly shaped aggregate, and part reflects the statistical uncertainty of the measurement due to background noise, but it is difficult to separate the two factors. 
Some particles show a periodic flux variability during the movie sequence likely due to rotation, but for most, the variability could be either due to rotation at a period ill sampled by the 6 minutes' duration and 40 s cadence of the movie, or to the statistical uncertainty of the background subtraction.

To constrain its size we use the average of all measured fluxes, $J_i$, for a given aggregate, $i$
\begin{equation}
s_i = \sqrt {J_i \frac{r_h^2 \Delta^2}{p \Phi(\alpha) I_\odot}},
\label{eq:size}
\end{equation}
where $p$, and $\Phi(\alpha)$ are the geometric albedo and phase function of the aggregate, $\alpha$ is the phase angle under which the particle was viewed, $r_h$ is the heliocentric distance in AU, $\Delta$ is the distance between aggregate and observer in metres, and $I_\odot =$ 1.5650 W m$^{-2}$ nm$^{-1}$ is the solar flux in the NAC F22 filter at 1 AU. $J_i$ is in units of W m$^{-2}$ nm$^{-1}$. We use $p=$0.065 and $\Phi(\alpha=90^\circ)$ = 0.02 \citep{fornasier-hasselmann2015}, assuming that the aggregates have similar optical properties as the nucleus. {In the absence of data describing the phase function of decimetre-sized aggregates, we use the nucleus albedo and phase function rather than a typical dust phase function, assuming that such large aggregates are more adequately described as ``mini-comets'' rather than as dust. The unknown albedo and phase function introduce an uncertainty of a factor of several into the derived sizes.
If the albedo were a factor of ten higher (e.g. due to a high ice content of the aggregates), the derived radius would be smaller by a factor of $\sim$3. We find that, if nucleus-like, the aggregates have equivalent radii between 1 and 18 cm (Figure~\ref{fig:sizes}). Aggregates having less than 9 cm radius are not well sampled, because for most of them it was difficult to determine their trajectories due to their high abundance and resulting low contrast. Such aggregates are present in the images in large numbers, but with decreasing size they blend into a diffuse background of granular texture rather than being visible as individual tracks. For particles with radii $>9$ cm we fit a power-law to the differential size distribution with an exponent of -4.0, but caution that the fit is valid only in the narrow size range between 9 and 18 cm and sensitively depends on the aperture sizes used for the photometry due to the comparatively small number of aggregates.
\begin{figure}
  \includegraphics[width=\columnwidth]{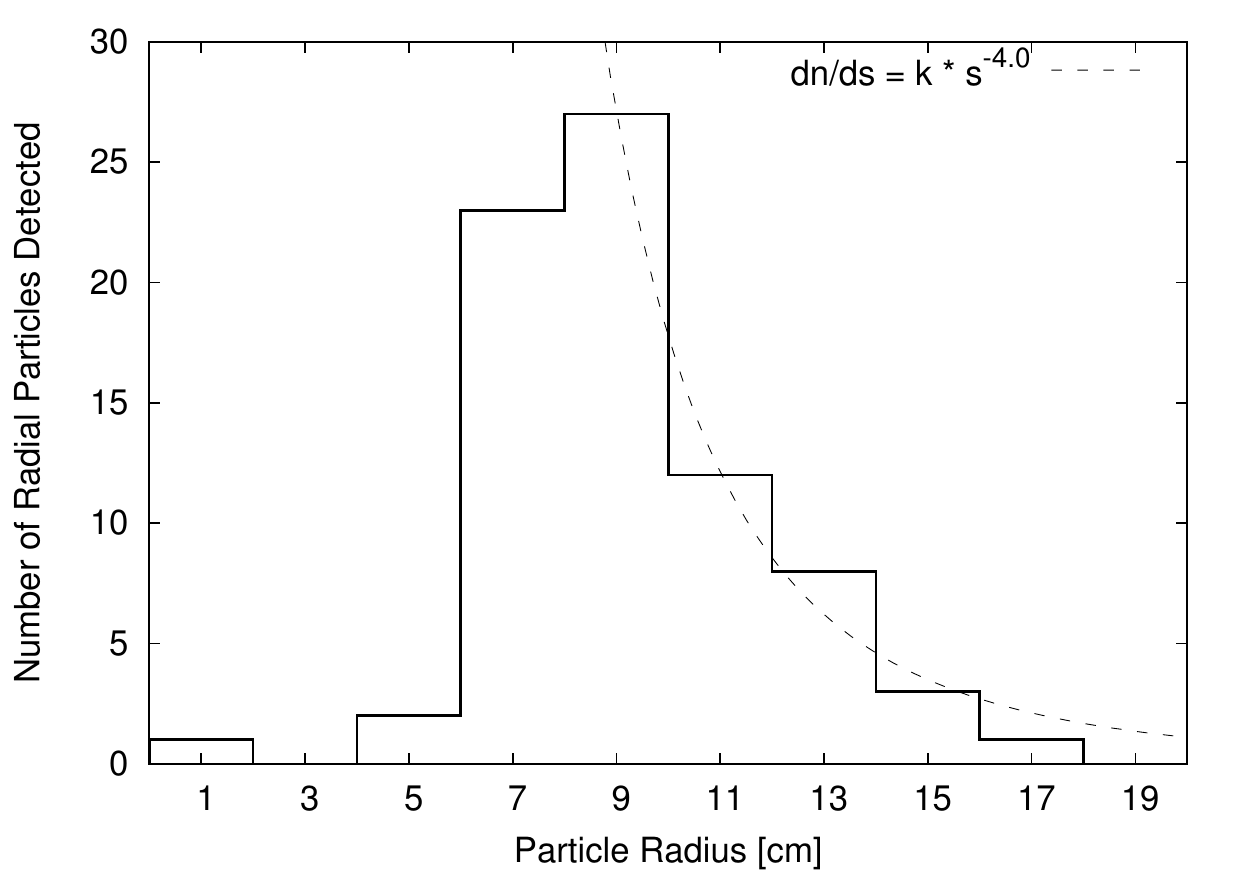}
    \caption{Histogram of the sizes of radially moving aggregates. The sampling of aggregates $<9$ cm in radius is incomplete due to difficulties detecting them, and the dashed line is a power-law fit to the data for radii $\geq 9$ cm.}
    \label{fig:sizes}
\end{figure}
Assuming a bulk density of 533 kg m$^{-3}$ (found for the nucleus by \citet{paetzold-andert2016}), the total mass contained in the 76 radially moving aggregates is 170 kg.

\section{Dynamical Analysis}
\label{sec:dynamics}
\subsection{Third Velocity Component}
\label{subsec:v3}
Our data only show us the aggregate motion projected to the image plane, and our derived velocities and accelerations are based on the assumption that the aggregates stay at a constant distance from the camera and that this distance is similar to that of the nucleus. We have no means to directly measure the aggregate motion parallel to the line of sight, but can constrain this component from the photometry. For each aggregate, the mean flux in the first and second half of the data differs by up to a factor 1.5. For constant aggregate size, this implies that the distance to the camera, $\Delta$, cannot have changed by more than a factor $\sqrt{1.5}$, approximately 20 km. 
\begin{figure}
  \includegraphics[width=\columnwidth]{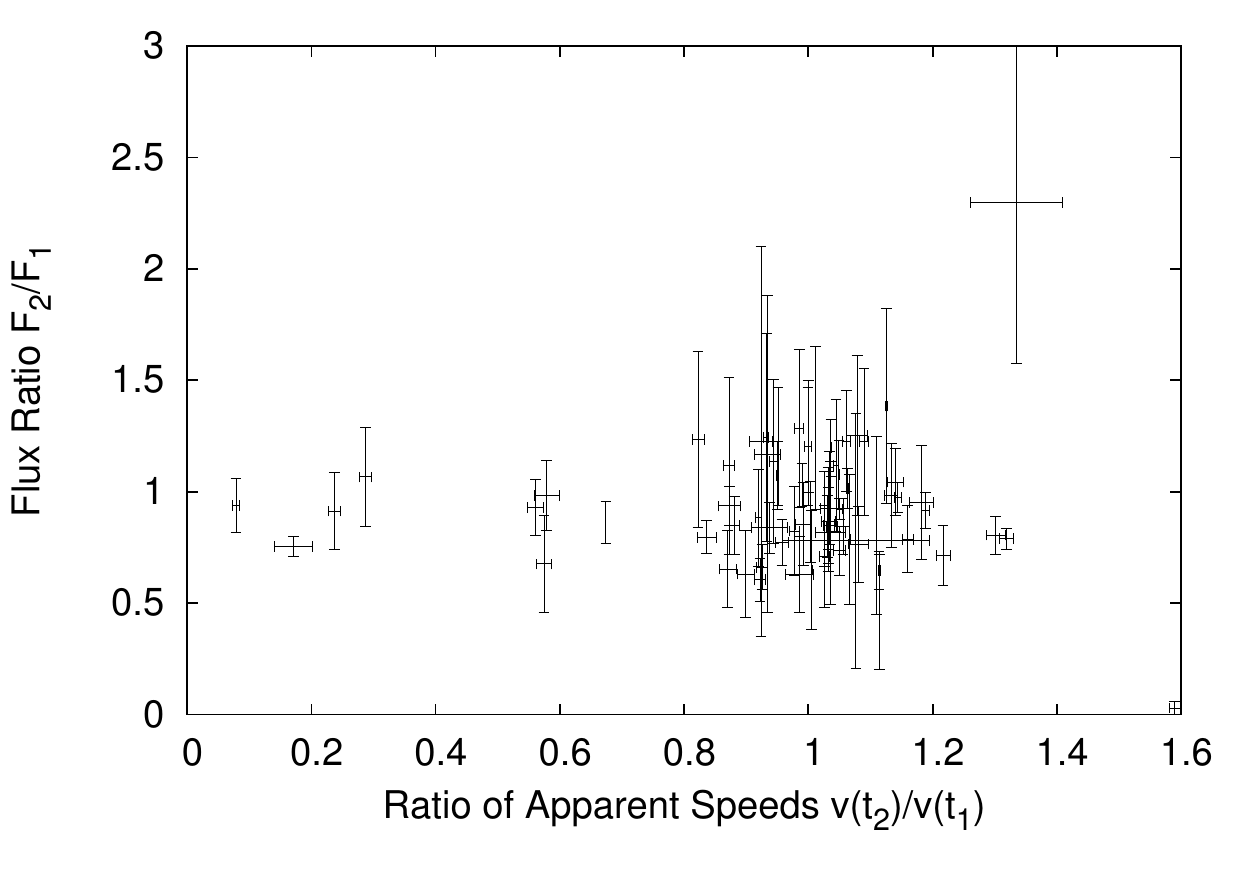}
    \caption{Ratio of the mean flux in the second and first halves of the images versus the ratio of the apparent speeds at the mean times corresponding to the two flux measurements. If the apparent acceleration were a projection effect, we would expect a correlation between these two quantities. Approximately one third of the aggregates have flux ratios $<$1 (including the error bars).
% The average flux ratio is 0.90 $\pm$ 0.04.
}
\label{fig:acc_vs_flux}
\end{figure}
Figure~\ref{fig:acc_vs_flux} does not show an obvious correlation between the relative change in flux with time and the change in apparent velocity. We conclude that the measured accelerations are real and not due to the distance dependence of the conversion from angular to physical pixel size. 

\subsection{Source Region}
To constrain the source region of the aggregates, we integrated their motion backwards in time, assuming a constant acceleration and using as initial conditions the positions and projected velocities derived from the movie frames. We use a coordinate system relative to the projected position of the nucleus centre and with the projected Sun direction parallel to the vertical image direction, which corresponds to constant nadir pointing. The reconstructed trajectories are shown in Figure~\ref{fig:traced_traj}. The upper left panel shows a reconstruction for approximately 40 minutes before the acquisition of the movie frame. The trajectories intersect in a region located approximately at 500 $< x <$ 1000 and 1300 $< y <$ 2000. We assume that this region of intersections corresponds to the projected source region at the time of emission, and that therefore the parts of the displayed trajectories corresponding to earlier times have no physical meaning, because at such times the aggregate in question would have been on the nucleus surface. The aggregates present in the images were likely emitted from the surface in the half hour before our observations.

The region of intersection seems to be elongated in the vertical direction, and the intersections seemingly correspond to different times: those trajectories crossing each other in their blue sections meet further down than those intersecting in their red parts. A conceivable explanation is that the source region of the aggregates moved with time upwards and slightly to the right, possibly due to the rotation of the nucleus. To better quantify the crossing times for different locations, we display the trajectories for 500 s sub-intervals and identify 8 regions where several trajectories of similar colour intersect. We assume that these regions indicate the projected locations of the source region(s) at the indicated times.
\begin{figure*}
  \includegraphics[width=1.9\columnwidth]{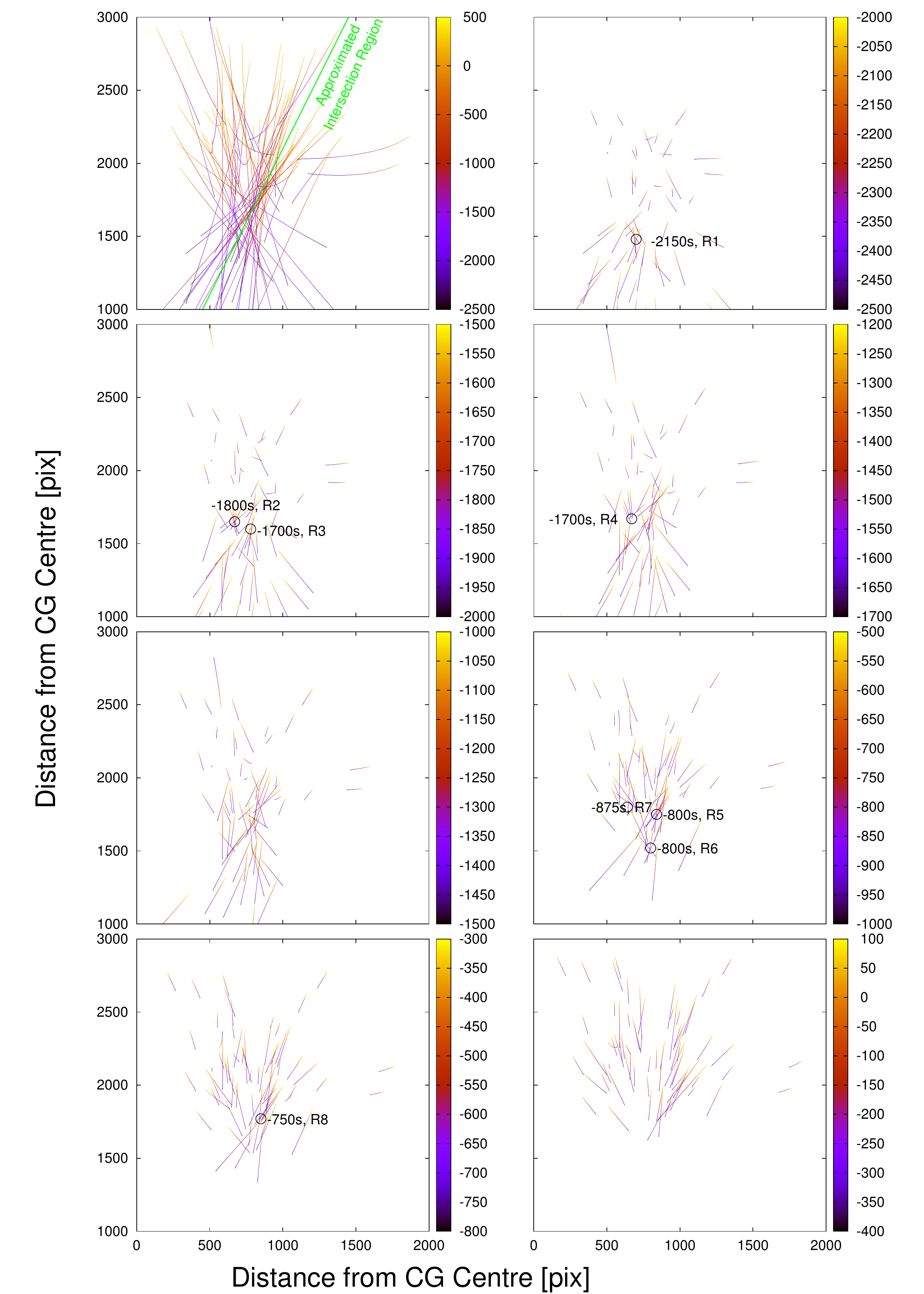}
    \caption{Reconstructed trajectories of aggregates. The motion was integrated backwards in time using as initial conditions the positions and velocities measured in the movie frames and assuming a constant acceleration. The colour code indicates time in seconds relative to UT 7:00. The upper left panel shows the time span from 6:18 to 7:08. The remaining panels shows the trajectories for 500 s sub-intervals and indicate regions (labeled R1 -- R8) where several trajectories seem to intersect at a given time. The green line in the top left panel shows an approximation of the intersection region by a line.
}
\label{fig:traced_traj}
\end{figure*}

To identify the position of the source region on the surface of the nucleus we calculate the intersection of the line of sight corresponding to each of the eight regions identified in Figure~\ref{fig:traced_traj} with a shape model of the nucleus \citep{jorda-gaskell2016} at the indicated time.  
\begin{figure}
  \includegraphics[width=\columnwidth]{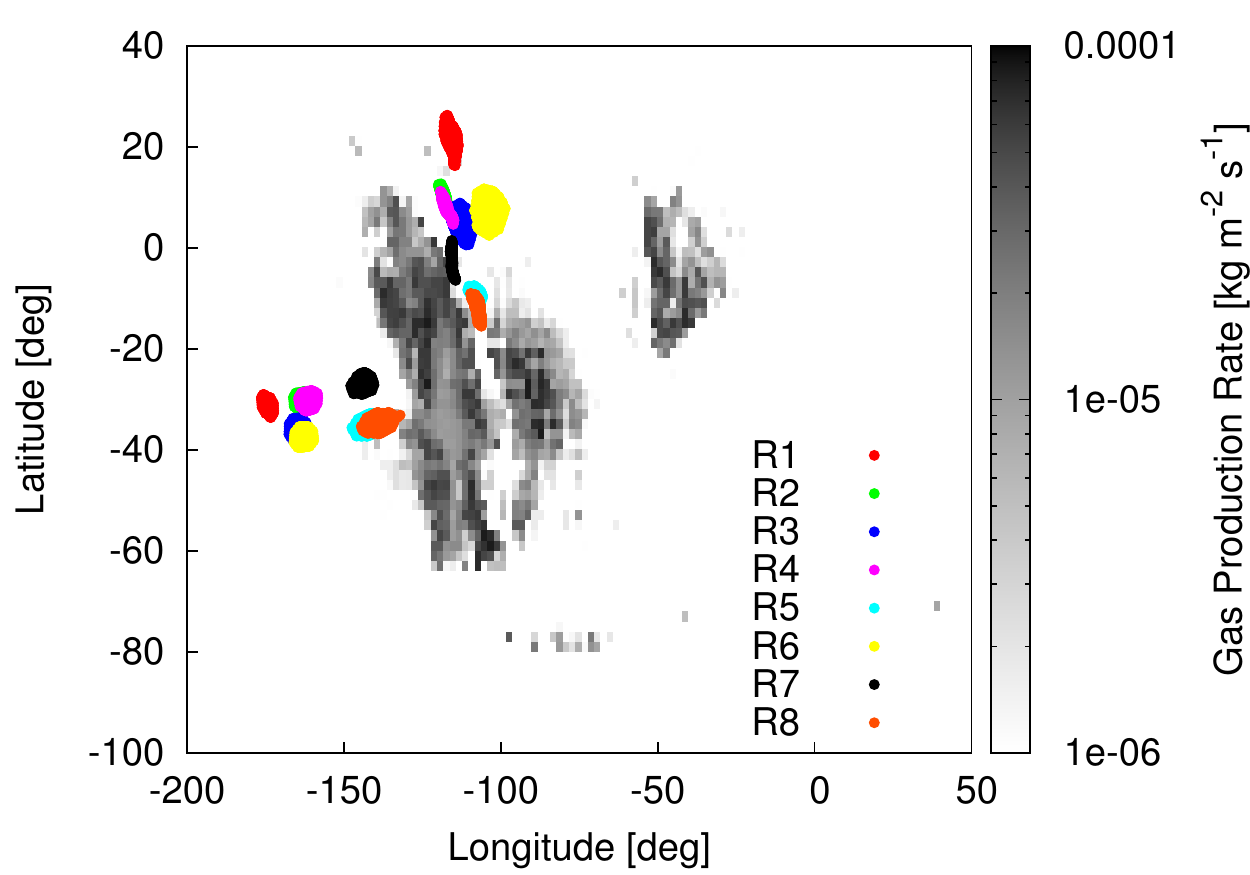}
    \caption{Location of the source regions identified in Figure~\ref{fig:traced_traj} on the surface of the comet. As each line of sight intersects the nucleus twice, two possible geographic source regions correspond to each projected source region. In the background, a map of the Hertz-Knudsen sublimation rate for a surface of pure water ice at 6:40 UT is shown. 
%The colour scale corresponds to gas production per shape model facet in kg s$^{-1}$.
}
\label{fig:source_reg}
\end{figure}
Each line of sight intersects the big nucleus lobe twice, such that each projected source region is associated with two possible geographic source regions. The geographic source regions are clustered in either the Khonsu region or in Seth.
The locations of the two regions on the surface of the comet are shown in Figure~\ref{fig:cg_viewer}.
\begin{figure}
  \includegraphics[width=\columnwidth]{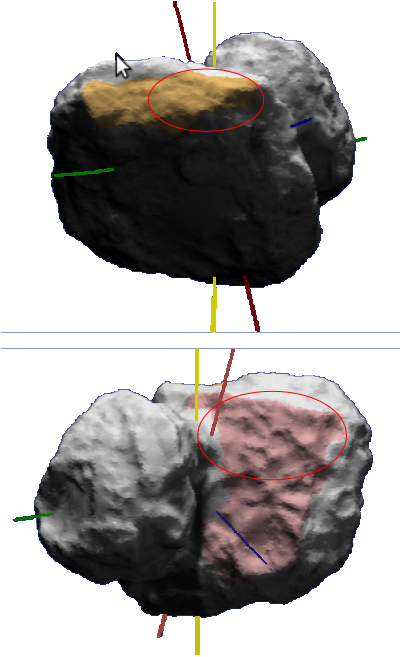}
    \caption{The nucleus of 67P as seen from Rosetta at the time of our observations (top panel) with the Khonsu region in colour, and (bottom) as an observer on the other side of the nucleus would have seen it with the Seth region coloured. The red ellipses roughly indicate the locations of our inferred source regions. The yellow line indicates the Sun-comet line, and the green, red, and blue lines represent the x-, y-, and z-axes in the standard reference frame of the comet. The z-axis coincides with the comet's rotation axis. The upper panel includes the South pole at the right edge of the big lobe, and the centre of the lower panel shows the North pole. The images were generated with the CG viewer tool of the OSIRIS team at https://planetgate.mps.mpg.de:8114.}
\label{fig:cg_viewer}
\end{figure}
There is no unique geographic region that matches all eight projected regions, but some regions match pair-wise (R2 and R4, R3 and R6, R5 and R8). It is possible that the geographic dispersion of the individual source regions reflects the presence of a larger area of activity, enveloping the individual regions. 

Alternatively we may have systematically overestimated the production age of the particles if the acceleration close to the surface was stronger than during the time of our observations. In that case we would have calculated the intersection between the line of sight and the nucleus shape for times too early, with a larger error for particles ejected later. This is supported by the apparent trend of ejection time with location on the surface: if the aggregates originated in the Khonsu region, the calculated ejection times increase from left to right, and in Seth they increase from top to bottom. We discuss the near-surface acceleration in more detail in Section~\ref{subsec:v0}. 

A third possible reason for the geographic dispersion of the source regions is that the activity location was indeed time-dependent, e.g. following the morning terminator in Khonsu. Figure~\ref{fig:source_reg} shows the gas production rate for a 11568-facet shape model assuming a surface of pure water ice. The local temperature is set to 50 K on the nightside and on the dayside modulated by the cosine of the solar incidence angle between a minimum value of 80 K and a maximum of 195 K, the latter given by the balance between incident solar flux, black-body thermal re-radiation and sublimation described by the Hertz-Knudsen sublimation rate.

%\begin{equation}
%T = \left( \frac{(1-A_B)L_\odot}{16 \pi \sigma \epsilon r_h^2} \right)^{1/4},
%\end{equation}
%where $A_B$ is the Bond albedo, $L_\odot$ is the luminosity of the Sun, $\sigma$ is the Stefan-Boltzmann constant and $\epsilon$ the emissivity at infrared wavelengths. 
%This simple temperature law ignores re-emission and sublimation cooling, and underestimates the temperature. On the other hand, the pure ice model
%overestimates the flux. Both effects balance each other and
%the end result is a good approximation of a more realistic model within a factor of $<$10. 

The spatial distribution of the modelled gas activity did not significantly change between 7:25 and 7:55 UT, such that this would not explain a displacement of the source region. Figure~\ref{fig:source_reg} shows that both reconstructed source areas are situated on the edges of the region with a high gas production. The Khonsu source region is located near the morning terminator, such that the ejection of our observed aggregates may relate to the local sunrise. The Seth source region is near the subsolar point where the maximum activity is expected, but the illumination in the region is shallow, resulting in a low gas production rate in our simple model. In reality, the gas production in Seth may be higher due to the thermal inertia of the surface.
We observe that the Seth source region is similar in shape to the gas producing region, and it may be that the two regions would match if we took into account an initially higher acceleration.

\subsection{Initial Velocities}
\label{subsec:v0}
Integrating the aggregate motion backwards in time also gives us the velocities of the aggregates while in the region of intersecting trajectories. To quantify the velocity in the intersection region, we approximate the intersection region by a line described by $y=200+2x$ (Figure~\ref{fig:traced_traj}), and the initial velocity by the velocity the aggregates had on crossing this line in our reconstruction based on a constant acceleration. 
\begin{figure}
  \includegraphics[width=\columnwidth]{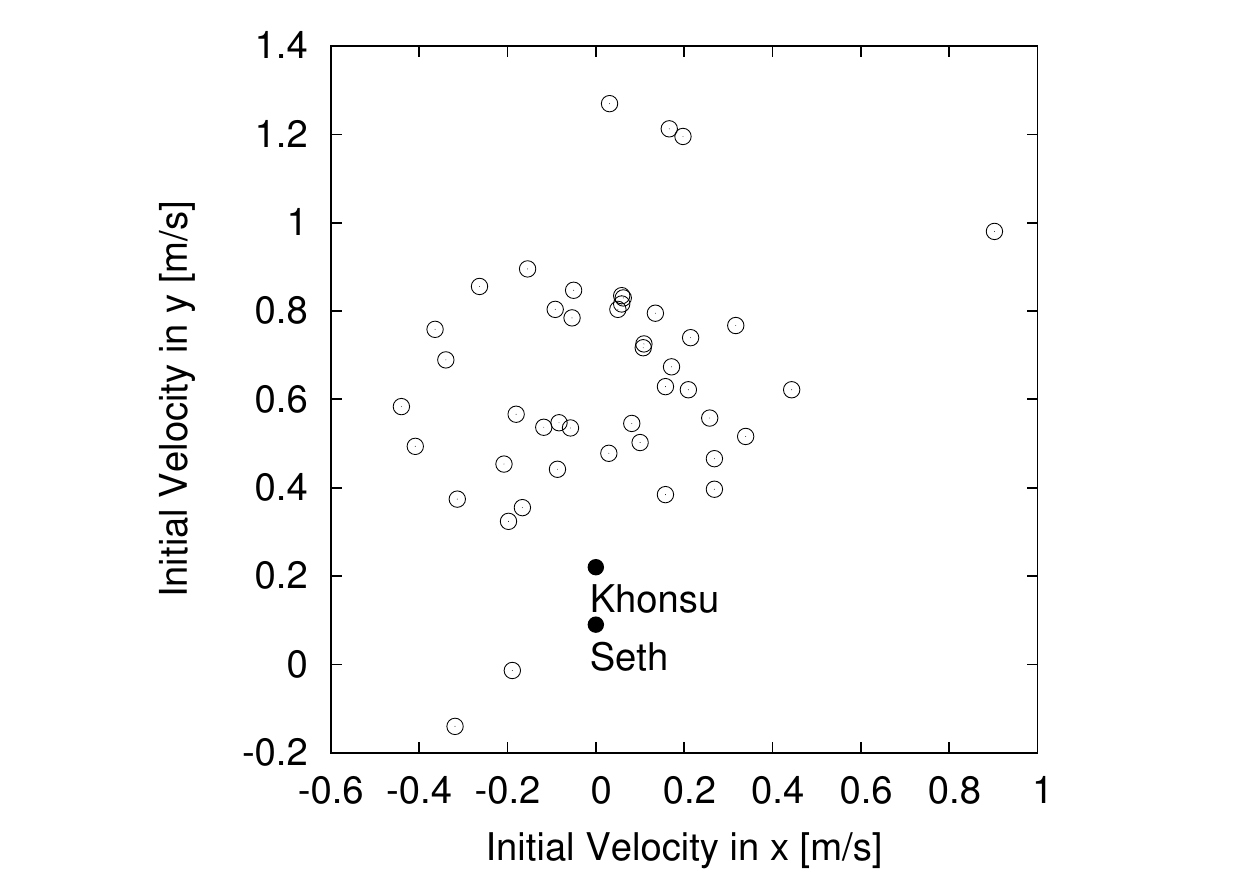}
    \caption{Open symbols: Initial velocity components of the aggregates reconstructed assuming a constant acceleration and the idealised source region shown in Figure~\ref{fig:traced_traj}. Filled symbols: The projected velocity components of the two candidate source regions due to the rotation of the nucleus.}
\label{fig:v_initial}
\end{figure}
Figure~\ref{fig:v_initial} shows the reconstructed initial velocity components in the image plane, and for comparison the rotational velocity of the surface in the two candidate source regions. The aggregate initial velocities are of order 0.4 to 0.9 m s$^{-1}$, compared to surface rotation speeds of 0.23 (Khonsu) and 0.09 m s$^{-1}$ (Seth). Part of the observed initial velocity is therefore due to the surface rotation, but in both cases, an additional acceleration during the first half hour of the aggregates' flight is required to explain the observed velocities. This can be either an increased acceleration at distances $<$1 km above the surface, or an initial ``kick'' the aggregates received during the process of decoupling from the surface, or both.

\subsection{Future Trajectories}
The local escape speed in the centre of our FOV and approximating the gravity field of the nucleus by that of a point mass of 9.982 $\times$ 10$^{12}$ kg \citep{paetzold-andert2016} is $v_{esc}$ = 0.55 m s$^{-2}$. About 10\% of the aggregates have projected velocities exceeding $v_{esc}$ and are subject to a positive vertical acceleration. These have a high chance of escaping from the gravity field of the nucleus and joining the comet's dust trail. 
Since we cannot measure the third velocity component perpendicular to the image plane, we cannot predict the future of the other 90\% of the aggregates. These have projected velocities below the local escape speed, and 50\% of them experience in addition a projected downward acceleration. It is possible that a fraction of these aggregates will eventually fall back to the surface as suggested by Figure~\ref{fig:invert}. Some aggregates may also enter bound orbits around the nucleus as suggested by \citet{richter-keller1995} and \citet{rotundi-sierks2015}.

\section{Accelerating Forces}
\label{sec:forces}
In the following we discuss the contribution to the aggregate acceleration of gravity, gas drag, and rocket force due to the sublimation of ice embedded in the aggregates. 

\subsection{Gravity}
We approximate the gravitational attraction of the nucleus by that of two point masses located at the centres of mass of the two nucleus lobes. The big lobe has its centre at (x,y,z) = (-0.673, 0.161, -0.040) km and a mass of 6.6 $\times$ 10$^{12}$kg, and the small lobe is at (1.523, -0.399, -0.219) km with a mass of 2.7 $\times$ 10$^{12}$ kg \citep{jorda-gaskell2016}. We calculated the components of the gravitational acceleration parallel to the image plane for varying distances from the spacecraft between 70 and 90 km, to account for our lack of precise knowledge on the distance of the aggregates. The projected gravity across our FOV is strongest ($\sim$ 10$^{-4}$ m s$^{-1}$) at spacecraft-comet distances between 86 and 88 km. The projected gravity vector in our FOV is inclined by up to 45$^\circ$ with respect to the vertical direction, and the apparent centre of gravity shifts from the left to the right edge of the FOV with increasing distance (Figure~\ref{fig:grav}).
\begin{figure}
  \includegraphics[width=\columnwidth]{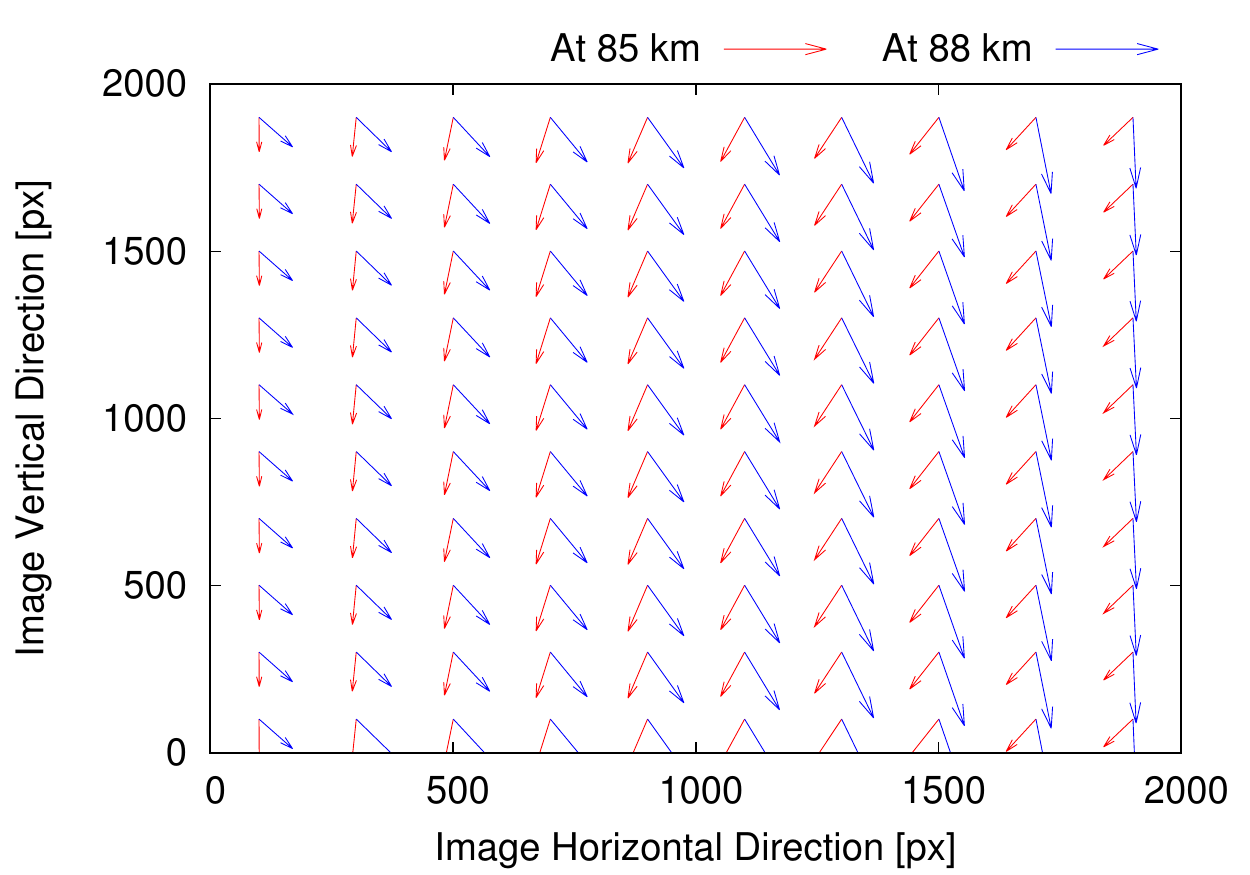}
    \caption{Projected gravity vector at distances of 85 and 88 km from the spacecraft. The magnitude of the acceleration is given by the length of the arrows in image pixels, multiplied by 0.5 $\times$ 10$^{-6}$ m s$^{-2}$.}
\label{fig:grav}
\end{figure}
The measured accelerations are up to one order of magnitude larger than gravity, suggesting that the aggregates must be subject to additional forces. 

\subsection{Rocket Force}
\label{subsec:rocket}
If the aggregates contain ice and sustain a temperature gradient across their surfaces, the asymmetric sublimation of ice will lead to a net force propelling the aggregates towards the direction of their coldest surface. The magnitude of this rocket-like acceleration is given by Eq.~12 in \citet{kelley-lindler2013}: 
\begin{equation}
a_{rocket} = \frac{3 \mu m_{\rm H} Z v_{th} f_{ice}}{4 \rho_p s},
\label{eq:a_rocket}
\end{equation}
where $\mu$ is the molecular weight of the sublimating ice, $m_{\rm H}$ is the mass of hydrogen, $Z$ is the sublimation rate in molec m$^{-2}$ s$^{-1}$, $v_{th}$ is the mean thermal expansion speed of the gas, $f_{ice}$, $\rho_p$, and $s$ are the ice fraction, bulk density and equivalent radius of the aggregate. We use $\mu$=18, $v_{th}$=500 m s$^{-1}$ \citep{kelley-lindler2013}, the bulk density and ice fraction of the nucleus, 500 kg m$^{-3}$ and 0.2 \citep{paetzold-andert2016}, and a sublimation rate of 4.1$\times$10$^{21}$ molec m$^{-2}$ s$^{-1}$, obtained from Eq.~14 of \citet{kelley-lindler2013} for compact dusty aggregates at the heliocentric distance of 2~AU. This gives us a rocket acceleration of $\sim$2$\times$10$^{-4}$ m s$^{-2}$, compatible with the magnitude of the measured acceleration (Figure~\ref{fig:acc}). However, the rocket effect cannot explain the acceleration to the positive vertical direction (towards the Sun). Depending on the rotation rate, axis orientation, and the thermal inertia of the aggregate, the warmest part of the surface can be at a solar zenith angle of up to 90$^{\circ}$, such that in our images obtained at a solar phase angle of 90$^\circ$, horizontal acceleration components in all directions are possible for randomly oriented aggregate spin axes, but the vertical acceleration must always be negative. We did not find a plausible thermal model for an aggregate to be warmer on its nightside than on the dayside. The rocket effect can therefore account for those aggregates subject to downward vertical acceleration, but not for the upward acceleration of the other 50\% of the aggregates.

If the aggregates had a significantly higher ice content, the albedo would be higher, resulting in a sublimation rate lower by a factor of 10 \citep{kelley-lindler2013}. Increasing the albedo by a factor 10 decreases the inferred aggregate size by a factor $\sim$3 for a given brightness. We expect that icy aggregates would have a density closer to 1000 kg m$^{-3}$, and an ice fraction of at least 1. Altogether, these various effects compensate, and we expect a similar magnitude of the rocket force also for aggregates with a higher ice content than the nucleus average assumed above.

An independent indication for on-going loss of embedded ice would be a decrease of aggregate brightness with time, due to decreasing albedo and/or size. Out of the 76 flux ratios shown in Fig.~\ref{fig:acc_vs_flux}, 30 are incompatible with a ratio of 1, and 27 out of these are $<$1, indicating that one third of the aggregates have faded with time. The fading was observed predominantly in those aggregates that we could trace for a sufficiently long time ($>$20 minutes), such that we cannot exclude an even higher fraction of fading aggregates, the detection of which was only prevented by our lack of a long-time measurement. If the brightness change was due to rotation, we would expect comparable numbers of aggregates getting brighter and fainter, and we have excluded in Section~\ref{subsec:v3} that the brightness evolution is due to changing distance. The fading of one third of the aggregates can, therefore, be an indication of on-going sublimation of ice.

\subsection{Gas Drag}
The drag force on a spherical dust grain of radius $s$ and bulk density $\rho_p$ embedded in a gas flow characterised by the velocity $v_{th}$ and gas mass density $\rho_g$ is given by 
\begin{equation}
a_{drag} = \frac{3}{8}  \frac{\rho_g}{\rho_p s} v_{th}^2 C_D,
\end{equation}
where $C_D$ is the drag coefficient of order unity.
%, ranging from 0.6 for a fluid dynamic interaction between gas and aggregate to 3 for a particulate flow (TBC reference).  
We extrapolate the gas density from the model presented in \citet{fougere-altwegg2016}. Based on Rosetta/ROSINA data, they find the global water production rate to depend on heliocentric distance as $Q_{H_2O} \propto r_h^{-4.2}$, and present a model coma for the date of 23 December 2014 at $r_h$ = 2.7 AU, such that we scale their calculated gas densities by a factor (2/2.7)$^{-4.2}$ = 3.5. Based on the model snapshots shown in \citet{fougere-altwegg2016} we estimate the gas density above the Khonsu region as 10$^{14}$ m$^{-3}$ and above Seth as 10$^{16}$ m$^{-3}$ in 2014, and correspondingly higher at the time of our observations. A mean expansion speed of 500 m s$^{-1}$ is compatible with \citet{fougere-altwegg2016}. We obtain gas drag accelerations of order 10$^{-8}$ m s$^{-2}$ above Khonsu and 10$^{-6}$ m s$^{-2}$ above Seth. Both are several orders of magnitude lower than our measured acceleration. However, there is considerable uncertainty in this estimate, because we do not exactly know the location above the surface and because we extrapolated from a model referring to a different season of the comet (southern winter as opposed to southern summer during our observation). Seasonal effects strongly influence the distribution of activity across the surface \citep{keller-mottola2015_erosion}. In addition, our images show that the aggregates are not ubiquitous but originate from a special region on the surface, such that it is conceivable that in this particular region the gas production was significantly above average. More sophisticated models of the gas flow at the time of our observations will be required to obtain a proper estimate of the gas drag force and to address the question whether the downward acceleration can as well be explained by gas, e.g. by a flow originating in the illuminated active area shown in Figure~\ref{fig:source_reg} and expanding laterally around the big lobe to the non-illumated area beyond the morning terminator.

\subsection{Synthesis}
\label{subsec:synthesis}
The simplest interpretation of the measured accelerations and their variability seems that all aggregates are subject to the combined forces of gravity, gas drag, and rocket effect. Gravity is similar (with variation in direction up to 45$^{\circ}$) for all aggregates. Also the gas drag will be similar for all aggregates: due to the narrow size range of the aggregates discussed here, the size-dependence of the drag force will introduce a variation within a factor of 2. In the simple model of a homogeneously outgassing surface region, the gas flow would be in the upward and to the lateral directions, as reflected in the fountain-like overall impression of the tracks in Figure~\ref{fig:tracks}.
%, such that from gas drag alone we expect accelerations only in the positive vertical and the two horizontal directions. 
We do not exclude that the complicated shape of the nucleus and a strongly inhomogeneous distribution of gas release from the surface can introduce a more complicated flow pattern, but for now try to explain our observations in the simplest possible way. 

Sticking to the simple gas model, the only explanation for the downward acceleration of 50\% of the aggregates is the rocket force due to sublimation of embedded ice in combination with a temperature gradient across the aggregates' surfaces. For rotating aggregates with radomly oriented spin axes, the rocket acceleration can be in all directions perpendicular to the solar direction, and will always point away from the Sun with a magnitude varying between 0 and $a_{rocket}$ (Eq.~\ref{eq:a_rocket}). Alternatively, the rocket force can vary between aggregates if these have different ice contents. Our observations do not allow us to distinguish between rotating grains with high thermal inertia and a sample of grains with variable ice content, or a mixture of both.
As our images were taken at a phase angle of $\sim$90$^{\circ}$, the solar direction coincides with the positive upward direction in the images.
%A sample of aggregates with randomly oriented spin axes will display downward vertical accelerations between a minimum (in the extreme zero) and the maximum acceleration, $a_{rocket}$, corresponding to a non-rotating aggregate or one with the spin-axis aligned with the solar direction. 

We expect to find a combination of gas drag and rocket for each aggregate. Those aggregates subject to the highest positive vertical acceleration would be those with the lowest vertical rocket acceleration. This gives a lower limit on the gas drag force of $a_{drag} \sim$4$\times$10$^{-4}$ m s$^{-2}$ from Figure~\ref{fig:acc}. This is consistent with the requirement that gas drag must exceed gravity ($\sim$ 10$^{-4}$ m s$^{-2}$) in order to lift the grains from the surface at all.
The aggregates with the strongest negative vertical acceleration experience the same gas drag as those accelerated upward, but in addition the strongest possible rocket acceleration, such that the strongest negative vertical acceleration corresponds to $a_{drag} + a_{rocket} \sim$ -3$\times$10$^{-4}$ m s$^{-2}$, and $a_{rocket} \sim$ -7$\times$10$^{-4}$ m s$^{-2}$, which is slightly above our derived value in Section~\ref{subsec:rocket}, but seems reasonable in light of the uncertainties of the sublimation rate, ice content, density, and size. 

In this interpretation of our data, the rocket force plays a decisive role in preventing the aggregates from escaping the gravity field of the nucleus. Aggregates having a lower ice content or ejected near the terminator would have a much higher probability of escape. 

The horizontal accelerations can be due both to drag from a laterally expanding gas flow and to the rocket force. In the latter case, aggregates travelling to the left of the source region would rotate in the opposite sense as those travelling to the right. From our present data we cannot infer the relative strength of the two effects.

\section{Conclusions}
\label{sec:conclusions}
We summarise our results, discuss their significance for understanding the activity of comet 67P, and outline open questions and future work.

\begin{itemize}
\item On 6 January 2016, we observed a fountain of aggregates having equivalent radii between 10 and 20 cm (assuming albedo and phase function of the nucleus) emerging from a localised area on the surface.

\item The likely source region is on the big nucleus lobe, either in Khonsu in the Southern hemisphere (at the time of ejection close to the morning terminator) or in Seth in the Northern hemisphere, close to local noon. 

\item The projected velocities of the aggregates are of order 1 m s$^{-1}$, comparable to the local escape speed from the nucleus.

\item The aggregates are subject to a constant acceleration over the 2h interval of our observations. 50\% of the aggregates are accelerated towards the projected nucleus direction, and 50\% away from it, towards the solar direction. The aggregates also receive a lateral acceleration to either side.

\item The measured accelerations are up to a factor 10 larger than gravity.

\item Correlation between horizontal velocity and acceleration suggests that the acceleration at work during our observations has been causal for the observed velocities. Candidate forces are (1) rocket force due to asymmetric sublimation of ice embedded in the aggregates, and (2) gas drag. Only gas drag can explain the upward acceleration, while the rocket force could explain the downward acceleration.

\item We observe a temporal brightness descrease in at least 1/3 of the aggregates, possibly indicative of the on-going sublimation of embedded ice.

\item Before the start of our observations, the aggregates must have either experienced a stronger upward acceleration (e.g. due to higher gas pressure close to the surface), or left the surface with an initial kick, or a combination of both.
\end{itemize}

In our interpretation, the measured accelerations result from the combined effects of gas drag pointing away from the surface and rocket force pointing away from the Sun and -- for the given geometry -- towards the surface. A random distribution of aggregate spin axes and/or a variable ice content can explain the observed range of the relative strengths of the two forces.

Alternatively, the lateral and downward acceleration can be induced by torques due to the interaction between rotating non-spherical, but not necessarily outgassing aggregates and the embedding gas flow \citep{fulle-ivanovski2015}. 

The aggregates discussed in this paper are likely to contribute to the material deposits covering parts of the northern hemisphere of the comet \citep{thomas-davidsson2015}. While the existence of material falling back to the surface has long been inferred from surface morphology, we here detect it in the process of ejection. Our conclusion that these aggregates were likely sublimating is consistent with the observation that the deposited material contains ice that contributes to the activity in these regions \citep{shi-hu2016, hu-shi2016_egu}. 

Our observations do not allow us to decide whether the aggregates correspond to the primordial building blocks of the comet or whether they are fragments of a crust formed during a later stage of the cometary evolution.

OSIRIS has carried out similar observations through significant parts of the Rosetta mission. A preliminary analysis shows that localised emission of large aggregates is common but not ubiquitous. Future analysis will reveal whether systematics exist regarding the ejection conditions and source location(s), and if this type of activity changed with orbital phase. Such a more global and systematic picture will help to distinguish between primordial pebbles and crust material.

\section*{Acknowledgements}
We thank the referee, William Reach, for his constructive suggestions that significantly helped to improve the manuscript. 
OSIRIS was built by a consortium of the Max-Planck-Institut f{\"u}r Sonnensystemforschung, G{\"o}ttingen, Germany, CISAS University of Padova, Italy, the Laboratoire d'Astrophysique de Marseille, France, the Instituto de Astrof\`{\i}sica de Andalucia, CSIC, Granada, Spain, the Research and Scientific Support Department of the European Space Agency, Noordwijk, The Netherlands, the Instituto Nacional de T\`ecnica Aeroespacial, Madrid, Spain, the Universidad Polit{\'e}echnica
de Madrid, Spain, the Department of Physics and Astronomy of Uppsala University, Sweden, and the Institut f{\"u}r Datentechnik und Kommunikationsnetze der Technischen Universit{\"a}t Braunschweig, Germany. The support of the national funding agencies of Germany (DLR), France(CNES), Italy(ASI), Spain(MEC), Sweden(SNSB), and the ESA Technical Directorate is gratefully acknowledged.

IRAF is distributed by the National Optical Astronomy Observatory, which is operated by the Association of Universities for Research in Astronomy (AURA) under a cooperative agreement with the National Science Foundation. 

%%%%%%%%%%%%%%%%%%%%%%%%%%%%%%%%%%%%%%%%%%%%%%%%%%



\bibliographystyle{mnras}
\bibliography{refs} % if your bibtex file is called example.bib

%\begin{thebibliography}{99}
%\end{thebibliography}

%%%%%%%%%%%%%%%%%%%%%%%%%%%%%%%%%%%%%%%%%%%%%%%%%%

%%%%%%%%%%%%%%%%% APPENDICES %%%%%%%%%%%%%%%%%%%%%

%\appendix



%%%%%%%%%%%%%%%%%%%%%%%%%%%%%%%%%%%%%%%%%%%%%%%%%%

% Don't change these lines
\bsp	% typesetting comment
\label{lastpage}
\end{document}